# Solution to an Anomaly in Internal Energy inside Nonextensive Statistical Mechanics

F. A R Navarro[(1)] and J. F. V. Flores[(2)]

**(1)** farnape@gmail.com, **(2)** jventounmsm@unmsm.edu

Education National University, National Mayor de San Marcos University,  Lima 14 -  Peru

## Abstract

Herein, in the context of third version of nonextensive statistical mechanics, a theory that generalizes the Boltzmann- Gibbs-Shannon's statistics, we display a solution for an anomaly found by calculating the internal energy for a composite $A+B$, of 2 spines ½, with additive Hamiltonian $H = H_A + H_B$. Specifically, the calculations of the internal energy in the full Hilbert space is different from the calculations done in the Hilbert subspaces, in other words, $U_{total}$ is different to $U_A + U_B$. We carry out analytical calculations. The results exactly indicate that the alternative method of matrices $E_A$ and $E_B$ is suitable for the calculations of the internal energy. Consequently, the matrix that holds the physical information of the system is $\rho^q$.

## I. Introduction

The motivation for this work are the researches that have been developed in the last 20 years with the nonextensive statistical mechanics [1], a theory developed by the Brazilian researcher C. Tsallis, which would be an alternative for Boltzmann-Gibbs-Shannon's statistics. The amount of scientific investigations with respect to this theory is large; nowadays, there are more than 2000 publications [2].Those studies have included various disciplines from quantum field theories to condensate matter physics. In this article, we analyzed a problem found in the internal energy (also called $q$-internal energy): the calculations of this thermodynamic potential, in different Hilbert spaces, produce result unequal. Previous researches for magnetization can be seen in the article *A Study on Composed Nonextensive Magnetic Systems* [3]. That paper display only computer simulations.

## II. Theoretical Frame

The construction of the nonextensive statistical mechanics begins by postulating the Tsallis entropy $S_q$ [4]:

$$S_q = k_B \frac{1 - \text{Tr}(\rho^q)}{q - 1}, \tag{1}$$

where $\rho^q$ is the density matrix $\rho$ powered to $q$, with $q$ the entropic index and $k_B$ the Boltzmann's constant. **Tr** symbolizes the trace operation over all states from $\rho^q$. In the limit, when $q$ tends to 1, we recovered the Boltzmann-Gibbs entropy,

$$S_1 = -k_B \text{Tr}(\rho \text{Ln} \rho), \tag{2}$$

next, we apply the maximum entropy method that considers the statistical mechanics as an inference process and was invented by the American Edward T. Jaynes in 1957 [5, 6]. And so, the following step is find out the probability distribution that maximizes the entropy of the Eq. (1); we want emphasize that the entropy is postulated without knowing the probability distribution *a priori*. For application of the method, we should take into account the constraints of unitary norm and a redefinition of the internal energy, that is,

$$\text{Tr}(\rho) = 1 \quad \text{and} \quad U_q = \frac{\text{Tr}(\rho^q H)}{\text{Tr}(\rho^q)}, \qquad (3)$$

$H$ is the Hamiltonian matrix. After maximization of $S_q$, we obtain the following probability distribution $p_i$:

$$p_i = \rho_{ii} = \frac{[1-(1-q)\beta'\varepsilon_i]^{\frac{1}{1-q}}}{Z_q}, \qquad (4)$$

$\rho_{ii}$ is the matrix element $ii$ of density matrix, $\beta' = \dfrac{1}{k_B T}$ is an energy parameter with $T$ the temperature, see the works [3, 7, 8]; $Z_q$ is the partition function,

$$Z_q = \text{Tr}[1-(1-q)\beta'\varepsilon_i]^{\frac{1}{1-q}}, \qquad (5)$$

For a compound system $A+B$, we also postulate the entropy of the complete system:

$$S_{A+B} = k_B \frac{1-\text{Tr}_{(A,B)}[\rho^q]}{q-1}, \qquad (6)$$

whereas for the entropies from subsystems $A$ and $B$, we again postulate them:

$$S_A = k_B \frac{1-\text{Tr}_A(\rho_A^q)}{q-1} \quad \text{and} \quad S_B = k_B \frac{1-\text{Tr}_B(\rho_B^q)}{q-1}, \qquad (7)$$

Nowadays, there are four versions of nonextensive statistical mechanics that are researched [9]; they all utilize the same entropy $S_q$ of Eq. (1). Nevertheless, the third version is the most accepted and widely investigated by the scientific community. This last version was devised by C. Tsallis, A. R. Plastino and R. S. Mendes in 1998 in the paper *The Role of Constraints within Generalized Nonextensive Statistics* [10].

## III. Procedure

We study a magnetic system of two sublattices $A$ and $B$. The sublattice $A$ has $N_A$ ions with spin ½, and the sublattice $B$ has $N_B$ ions also with spin ½. We utilize the Heisenberg model in the mean field approach. Hereafter, the Hamiltonian and other physical observables will be represented by operators, that is:

$$\hat{H}_{A+B} = \hat{H}_A + \hat{H}_B, \qquad (8)$$

with

$$\hat{H}_A = g\mu_h B_A \sum_i^N \hat{S}_{A,i}^z \quad \text{and} \quad \hat{H}_B = g\mu_h B_B \sum_i^N \hat{S}_{B,i}^z, \qquad (9)$$

$g$ is the gyromagnetic factor, $\mu_h$ is the Bohr magneton, for didactic reasons we changed the usual symbol $\mu_B$; $\hat{S}_{A,i}^z$ and $\hat{S}_{B,i}^z$ spin operators, $A(B)$ indicates respective sublattice and the subscript $i$, the $i$-th ion; $B_A$ and $B_B$ are the called effective magnetic fields and they are defined by:

$$B_A = B_0 + \lambda_A M_A + \lambda_{AB} M_B \quad \text{and} \quad B_B = B_0 + \lambda_B M_B + \lambda_{AB} M_A, \qquad (10)$$

$B_0$ is the external magnetic field; $\lambda_A$ and $\lambda_B$ are the intralattice coupling parameters and $\lambda_{AB}$ the interlattice coupling parameter. By considering that spines are independent the Eq. (9) is expressed in this way:

$$\hat{H}_A = g\mu_h B_A N_A \hat{S}_A^z \quad \text{and} \quad \hat{H}_B = g\mu_h B_B N_B \hat{S}_B^z. \qquad (11)$$

In the following subsections we show calculations to finding the composite internal energies.

## 3.1 Calculation of the Internal Energies in Full Hilbert Space 4x4

In this subsection we apply the nonextensive statistical mechanics for researching the magnetic system represented by the Eqs. (8) and (9). For simplifying the calculations, we allow for $N_A=N_B=N$. In full Hilbert space, the internal energies have the general formulas:

$$U_A = \frac{\text{Tr}_{(A,B)}(\hat{\rho}^q \hat{H}_A)}{\text{Tr}_{(A,B)}(\hat{\rho}^q)} \quad \text{and} \quad U_B = \frac{\text{Tr}_{(A,B)}(\hat{\rho}^q \hat{H}_A)}{\text{Tr}_{(A,B)}(\hat{\rho}^q)}, \quad (12)$$

next, for our specific system of 2 spines ½, in a Hilbert space 4x4, the respective internal energies take the form:

$$U_A = -Ng\mu_h B_A \frac{\text{Tr}_{(A,B)}(\rho^q \hat{S}_A^z)}{\text{Tr}_{(A,B)}(\rho^q)} \quad \text{and} \quad U_B = -Ng\mu_h B_B \frac{\text{Tr}_{(A,B)}(\rho^q \hat{S}_B^z)}{\text{Tr}_{(A,B)}(\rho^q)}. \quad (13)$$

In order to calculate these parameters, we will need the density matrix operator, as well as the spin operators. Firstly, we obtain the density matrix operator from the Eq. (4):

$$\rho = \frac{\left\{1-(1-q)\frac{2N\mu_h}{K_B T}(\hat{S}_A^z B_A + \hat{S}_B^z B_B)\right\}^{\frac{1}{1-q}}}{Z_q}, \quad (14)$$

which we power to $q$:

$$\rho^q = \frac{\left\{1-(1-q)\frac{2N\mu_h}{k_B T}(S_A^z B_A + S_B^z B_B)\right\}^{\frac{q}{1-q}}}{Z_q^q}. \quad (15)$$

On the other hand, the dimensionless spin operators in the full Hilbert space 4x4 are (the constant $\hbar$ is included in the Bohr magneton):

$$\hat{S}_A^z = \frac{\left\{|++\rangle\langle++| + |+-\rangle\langle+-| - |-+\rangle\langle-+| - |--\rangle\langle--|\right\}}{2} \quad \text{and}$$

$$\hat{S}_B^z = \frac{\left\{|++\rangle\langle++| - |+-\rangle\langle+-| + |-+\rangle\langle-+| - |--\rangle\langle--|\right\}}{2} \quad (16)$$

where $|++\rangle\langle++|$, $|+-\rangle\langle+-|$, $|-+\rangle\langle-+|$ and $|--\rangle\langle--|$ are external operators that are constructed by bras and kets, which represent quantum states of the complete system. The bras are $\langle++|$, $\langle+-|$, $\langle-+|$ and $\langle--|$; the respective kets are $|++\rangle$, $|+-\rangle$, $|-+\rangle$ and $|--\rangle$. Now, we can obtain the density matrix elements of $\hat{\rho}$,

$$\rho_{++} = \langle ++|\hat{\rho}|++\rangle = \alpha_1/Z_q, \qquad \rho_{+-} = \langle +-|\hat{\rho}|+-\rangle = \alpha_2/Z_q,$$

$$\rho_{-+} = \langle -+|\hat{\rho}|-+\rangle = \alpha_3/Z_q \qquad \rho_{--} = \langle --|\hat{\rho}|--\rangle = \alpha_4/Z_q$$
and (17)

for practical reasons, we have introduced the following parameters:

$$\alpha_1 = \left[1-(1-q)\frac{2N\mu_h(B_A+B_B)}{k_BT}\right]^{\frac{1}{1-q}}, \qquad \alpha_2 = \left[1-(1-q)\frac{2N\mu_h(B_A-B_B)}{k_BT}\right]^{\frac{1}{1-q}}$$

$$\alpha_3 = \left[1-(1-q)\frac{2N\mu_h(-B_A+B_B)}{k_BT}\right]^{\frac{1}{1-q}} \text{ and } \alpha_4 = \left[1+(1-q)\frac{2N\mu_h(B_A+B_B)}{k_BT}\right]^{\frac{1}{1-q}}; \qquad (18)$$

so that we also can get the matrix elements of $\hat{\rho}^q$:

$$\rho^q_{++} = \langle ++|\hat{\rho}^q|++\rangle = \alpha_1^q/Z_q^q, \qquad \rho^q_{+-} = \langle +-|\hat{\rho}^q|+-\rangle = \alpha_2^q/Z_q^q,$$

$$\rho^q_{-+} = \langle -+|\hat{\rho}^q|-+\rangle = \alpha_3^q/Z_q^q \qquad \rho^q_{--} = \langle --|\hat{\rho}^q|--\rangle = \alpha_4^q/Z_q^q.$$
and (19)

Therefore, by using the external operators, the density matrix operator $\hat{\rho}$ can be rrewritten as follows:

$$\hat{\rho} = \left\{\alpha_1|++\rangle\langle++| + \alpha_2|+-\rangle\langle+-| + \alpha_3|-+\rangle\langle-+| + \alpha_4|--\rangle\langle--|\right\}/Z_q, \qquad (20)$$

and, analogously, for $\hat{\rho}^q$ we have:

$$\hat{\rho}^q = \left\{\alpha_1^q|++\rangle\langle++| + \alpha_2^q|+-\rangle\langle+-| + \alpha_3^q|-+\rangle\langle-+| + \alpha_4^q|--\rangle\langle--|\right\}/Z_q^q, \qquad (21)$$

next, replacing Eqs. (16) and (21) into the Eq. (13), we obtain the internal energies in the full Hilbert space 4x4:

$$U_A = -Ng\mu_hB_A\frac{\alpha_1^q+\alpha_2^q-\alpha_3^q-\alpha_4^q}{\alpha_1^q+\alpha_2^q+\alpha_3^q+\alpha_4^q} \text{ and } U_B = -Ng\mu_hB_B\frac{\alpha_1^q+\alpha_3^q-\alpha_2^q-\alpha_4^q}{\alpha_1^q+\alpha_2^q+\alpha_3^q+\alpha_4^q}, \qquad (22)$$

and so, we also get the total internal energy:

$$U_{A+B} = -Ng\mu_h\frac{B_A\{\alpha_1^q+\alpha_2^q-\alpha_3^q-\alpha_4^q\}+B_B\{\alpha_1^q+\alpha_3^q-\alpha_2^q-\alpha_4^q\}}{\alpha_1^q+\alpha_2^q+\alpha_3^q+\alpha_4^q}. \qquad (23)$$

In following two subsections, by using two procedures, these results will be compared with calculations in Hilbert subspaces 2x2.

## 3.2 Calculation of the Internal Energies in Hilbert Subspaces 2x2, by Using the Formulas of the Third Version

Now, we calculate $U_A$ and $U_B$ in Hilbert subspaces 2x2. In principle, the results must be identical with the outcomes obtained in the Hilbert space 4x4. In the Hilbert subspaces 2x2 the internal energies for the system of the Eqs. (8) and (9) are defined as:

$$U_A = -Ng\mu_h B_A \frac{\text{Tr}_A(\hat{\rho}_A^q \hat{s}_A^z)}{\text{Tr}_A(\hat{\rho}_A^q)} \quad \text{and} \quad U_B = -Ng\mu_h B_B \frac{\text{Tr}_B(\hat{\rho}_B^q \hat{s}_B^z)}{\text{Tr}_B(\hat{\rho}_B^q)}, \tag{24}$$

in order to calculate these thermodynamic parameters, we take the partial trace over $\hat{\rho}$ in the Eq. (20), so we obtain the respective partial matrices $\hat{\rho}_A$ and $\hat{\rho}_B$:

$$\hat{\rho}_A = \frac{\{(\alpha_1 + \alpha_2)|+\rangle\langle+| + (\alpha_3 + \alpha_4)|-\rangle\langle-|\}}{Z_q} \quad \text{and} \quad \hat{\rho}_B = \frac{\{(\alpha_1 + \alpha_3)|+\rangle\langle+| + (\alpha_2 + \alpha_4)|-\rangle\langle-|\}}{Z_q}, \tag{25}$$

now, we elevate to power $q$ and obtain

$$\hat{\rho}_A^q = \frac{\{(\alpha_1 + \alpha_2)^q|+\rangle\langle+| + (\alpha_3 + \alpha_4)^q|-\rangle\langle-|\}}{Z_q^q} \quad \text{and} \quad \hat{\rho}_B^q = \frac{\{(\alpha_1 + \alpha_3)^q|+\rangle\langle+| + (\alpha_2 + \alpha_4)^q|-\rangle\langle-|\}}{Z_q^q}. \tag{26}$$

For our calculation, we also know that the spin operators in the Hilbert subspaces 2x2 are:

$$\hat{s}_A^z = \frac{\{|+\rangle\langle+| - |-\rangle\langle-|\}}{2} \quad \text{and} \quad \hat{s}_B^z = \frac{\{|+\rangle\langle+| - |-\rangle\langle-|\}}{2}; \tag{27}$$

in this manner, by replacing Eqs. (26) and (27) into Eq. (24), we get the following result in the respective Hilbert spaces 2x2:

$$U_A = -Ng\mu_h B_A \frac{(\alpha_1 + \alpha_2)^q - (\alpha_3 + \alpha_4)^q}{(\alpha_1 + \alpha_2)^q + (\alpha_3 + \alpha_4)^q} \quad \text{and} \quad U_B = -N\mu_h B_A \frac{(\alpha_1 + \alpha_3)^q - (\alpha_2 + \alpha_4)^q}{(\alpha_1 + \alpha_2)^q + (\alpha_3 + \alpha_4)^q}, \tag{28}$$

so that the total internal energy is:

$$U_{A+B} = -Ng\mu_h \frac{B_A\{(\alpha_1 + \alpha_2)^q - (\alpha_3 + \alpha_4)^q\} + B_B\{(\alpha_1 + \alpha_3)^q - (\alpha_2 + \alpha_4)^q\}}{(\alpha_1 + \alpha_2)^q + (\alpha_3 + \alpha_4)^q}, \tag{29}$$

we notice that these outcomes are different from the results got in the full Hilbert space 4x4, Eqs. (22) and (23). This is interpreted to be an anomalous result but we solve it in the next subsection.

## 3.3 The Matrices $E_A$ and $E_B$ for the Calculation of the Internal Energies, in the Hilbert Subspaces 2x2

For the magnetic system represented by the Eqs. (8) and (9), we introduce a procedure that was known in [7]. Herein, for didactic reasons, we called $E_A$ and $E_B$ to the matrices $\rho_{A,q}$ and $\rho_{B,q}$ of the reference [7]. Therefore, in this context, we define the internal energies as:

$$U_A = -Ng\mu_h B_A \frac{\text{Tr}_A(\hat{E}_A \hat{s}_A^z)}{\text{Tr}_A(\hat{E}_A)} \quad \text{and} \quad U_B = -Ng\mu_h B_B \frac{\text{Tr}_B(\hat{E}_B \hat{s}_B^z)}{\text{Tr}_B(\hat{E}_B)} \tag{30}$$

where partial matrices operators $\hat{E}_A$ and $\hat{E}_B$ are defined as:

$$\hat{E}_A = \text{Tr}_B(\hat{\rho}^q) \quad \text{and} \quad \hat{E}_B = \text{Tr}_A(\hat{\rho}^q), \tag{31}$$

then, we carry out the partial trace over $\hat{\rho}^q$ in the Eq. (21):

$$\hat{E}_A = \frac{\{(\alpha_1^q + \alpha_2^q)|++\rangle\langle++| + (\alpha_3^q + \alpha_4^q)|--\rangle\langle--|\}}{Z_q^q} \quad \text{and}$$

$$\hat{E}_B = \frac{\{(\alpha_1^q + \alpha_3^q)|++\rangle\langle++| + (\alpha_2^q + \alpha_4^q)|--\rangle\langle--|\}}{Z_q^q}, \tag{32}$$

*we must emphasize that these matrices no longer need to be powered q*. Therefore, by replacing Eqs. (27) and (32) into Eq. (30), we found that the employment of the matrices $E_A$ and $E_B$ produces the following energies, in the Hilbert subspaces 2x2:

$$U_A = -Ng\mu_h B_A \frac{\alpha_1^q + \alpha_2^q - \alpha_3^q - \alpha_4^q}{\alpha_1^q + \alpha_2^q + \alpha_3^q + \alpha_4^q} \quad \text{and} \quad U_B = -Ng\mu_h B_B \frac{\alpha_1^q + \alpha_2^q - \alpha_3^q - \alpha_4^q}{\alpha_1^q + \alpha_2^q + \alpha_3^q + \alpha_4^q}. \tag{33}$$

Hence, it follows that the total energy of system $A+B$ is:

$$U_{A+B} = -Ng\mu_h \frac{B_A\{\alpha_1^q + \alpha_2^q - \alpha_3^q - \alpha_4^q\} + B_B\{\alpha_1^q + \alpha_3^q - \alpha_2^q - \alpha_4^q\}}{\alpha_1^q + \alpha_2^q + \alpha_3^q + \alpha_4^q}; \tag{34}$$

These last results agree exactly with the Eqs. (22) and (23) that were calculated in the full Hilbert space 4x4. This fact shows clearly that partial matrices $E_A$ and $E_B$ are the adequate ones for the calculation of the internal energy.

## IV. CONCLUSIONS

By using a magnetic system of 2 spines ½ as a specific example, we have shown explicitly that happen an anomaly by calculating the internal energy of a composite $A+B$, when the formulas from third version of nonextensive statistical mechanics are utilized, in the Hilbert subspaces. Also, we demonstrated that the solution to that problem is introducing matrices $E_A$ and $E_B$, these matrices contain the physical information of the subsystems. Therefore, the operation of partial trace should be taken on the matrix $\rho^q$ but not on the matrix $\rho$. Finally, we want stress that the proposal of the method with matrices $E_A$ and $E_B$ is confirmed by an exact analytical way.

## VI. BIBLIOGRAFÍA